\documentclass[aps,pra,letterpaper,superscriptaddress,twocolumn,showpacs,floatfix,10pt]{revtex4-1}
\usepackage{amsfonts}
\usepackage{amsmath}
\usepackage{amssymb}
\usepackage{graphicx}
\usepackage{epstopdf}
\usepackage{epsfig}
\usepackage{color}
\usepackage{mathtools}

\usepackage{hyperref}

\newcommand{\eket}[1]{ \left \vert #1 \right \rangle}
\newcommand{\ebra}[1]{ \left \langle #1 \right \vert}

\newcommand{\plaqa}{
 {\mathchoice
  {\includegraphics[height=1.6ex]{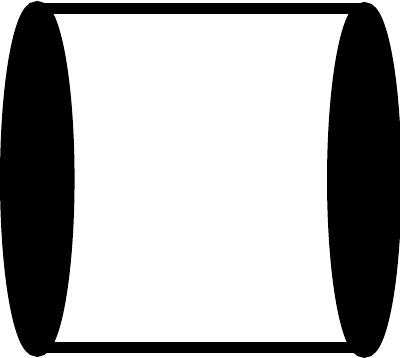}}
  {\includegraphics[height=1.6ex]{plaqa}}
  {\includegraphics[height=1.2ex]{plaqa}}
  {\includegraphics[height=0.9ex]{plaqa}}
 }
}
\newcommand{\plaqb}{
 {\mathchoice
  {\includegraphics[height=1.6ex]{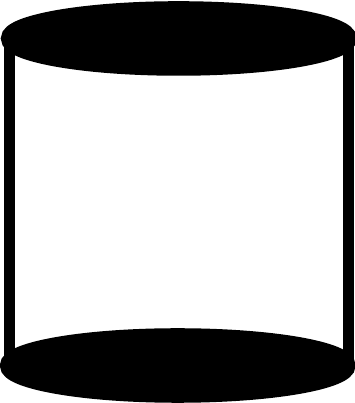}}
  {\includegraphics[height=1.6ex]{plaqb}}
  {\includegraphics[height=1.2ex]{plaqb}}
  {\includegraphics[height=0.9ex]{plaqb}}
 }
}

\begin{document}


\title{Tensor renormalization group approach to classical dimer models}

\author{Krishanu Roychowdhury }
\affiliation{Max-Planck-Institut f$\ddot{u}$r Physik komplexer Systeme, 01187 Dresden, Germany}    

\author{Ching-Yu Huang}
\affiliation{Max-Planck-Institut f$\ddot{u}$r Physik komplexer Systeme, 01187 Dresden, Germany}   
 \affiliation{C. N. Yang Institute for Theoretical Physics and Department of Physics and Astronomy, State University of New York at Stony Brook, NY 11794-3840, United States}


\vfill
\begin{abstract}

We analyze classical dimer models on the square and triangular lattice using a tensor network representation of the dimers.
The correlation functions are numerically calculated using the recently developed ``Tensor renormalization group'' (TRG) technique.
The partition function for the dimer problem can be calculated exactly by the Pfaffian method which is used here as a platform for comparing the numerical results.
TRG turns out to be a powerful tool for describing gapped systems with exponentially decaying correlations very efficiently due to its fast convergence.
This is the case for the dimer model on the triangular lattice.
However, the convergence becomes very slow and unstable in case of the square lattice where the model has algebraically decaying correlations.
We highlight these aspects with numerical simulations and critically appraise the robustness of TRG approach by contrasting the results for small and large system sizes against the exact calculations.
Furthermore, we benchmark our TRG results with classical Monte Carlo (MC) method.

\end{abstract}

\maketitle


\section{Introduction}


The problem of covering a planar graph with dimers subject to certain hard-core constraints, has attracted considerable attention in various solvable forms in the field of statistical mechanics as well as other branches of physics \cite{okounkov2006quantum}.
These classical models, although emerged from a pure mathematical origin, have been utilized to explain many physical phenomena, such as absorption of diatomic molecules \cite{roberts1939some}, high-temperature superconductivity \cite{RK1988}, frustrated magnetism \cite{diep2004frustrated, lacroix2011introduction} etc..
In one of the seminal papers on statistical mechanics, Fisher first pointed out that the frustrated Ising model on planar lattices can be mapped to classical dimer model on the dual lattices \cite{Fisher1961} corresponding to the same degenerate ground state manifold.
The zero-temperature frustrated Ising-spin model on triangular lattice, for example, can be translated to the dimer model on the dual honeycomb lattice where the dimers essentially represent the frustrated honeycomb links.
An important step towards solving this kind of problem and calculating the correlations of relevant observables was the introduction of the Pfaffian techniques \cite{ Kasteleyn1961, Fisher1961}, which turned out to be very useful for the simplest dimer problem on square lattice where a critical phase can be realized with algebraic dimer correlation \cite{alet2005interacting}.
Later a thermal phase transition of Kosterlitz-Thouless type was observed in an interacting dimer problem on square lattice separating the high temperature critical phase (algebraic correlation) and a low temperature crystalline dimer phase, and characterized in terms of the field theory of height variables \cite{PhysRevE.74.041124}.

A quantum version of the dimer model at zero temperature can also host many exotic phases particularly in context of realizing liquid states with topological imprints.
The effective Hamiltonian describing the quantum model, as introduced by Rokhsar and Kivelson \cite{RK1988} for a square lattice, comprises of two mutually competing elements serving as the potential and the kinetic term with amplitude $v$ and $t$ respectively and reads as follows,
\begin{align}
\mathcal{H}_{\text{eff}} = -t \sum_p \Bigl( \eket{\plaqa}\ebra{\plaqb} + \mathrm{h.c.} \Bigr) + v \sum_p \Bigl( \eket{\plaqa}\ebra{\plaqa} +\eket{\plaqb}\ebra{\plaqb} \Bigr). 
\label{Hrk}
\end{align}
where the sum over $p$ is performed over all square plaquettes. 
Similar Hamiltonian exists for the triangular lattice dimer model also though the phase diagram is distinct from the square lattice in several aspects \cite{Moessner2001}.
However, as a common feature to both of them, there exists a particular point at $v=t$, known as Rokhsar-Kivelson (RK) point, where the topologically degenerate ground state attains exactly zero energy in the form of an equal weight superposition of all possible configurations in a given winding parity sector.
All the correlations, at this special point, can be computed classically since the particular form of the ground state wavefunction maps the quantum model to an infinite temperature classical dimer model with equal weightage to all configurations.

Both the classical and the quantum model have been extensively studied using several numerical techniques such as classical Monte Carlo (MC) \cite{PhysRevE.74.041124, Fendley2002}, quantum Monte Carlo \cite{Moessner2001, ralko2005zero} etc..
Besides these traditional methods, a whole new series of simulation algorithms proliferated in last few years aiming at the efficient representation of many-particle quantum states based on the ideas motivated mostly from the subjects of quantum chemistry and quantum information theory \cite{verstraete2002four}.
The family contains members such as projected entangled-pair states (PEPS) \cite{peps_Cirac2004}, tree tensor state \cite{tree_Vidal2006}, and multiscale renormalization ansatz \cite{Vidal2007, Vidal2008}, all based on representing the fundamental degrees of freedom by tensors defined on the lattice.
For example, the tensor product representations have recently been used  successfully for simulating some of the classical lattice models and computing  the partition function along with other physical quantities in presence of local interactions \cite{trg07-Levin, zhao2010renormalization}.
In Ref.\cite{trg07-Levin}, the authors proposed a real space renormalization of the tensors, coined as tensor renormalization group (TRG), much in the spirit of block spin scheme used in the coarse graining of usual renormalization group (RG) approach and estimated some of the critical exponents of the square lattice Ising model with good accuracy. 
On the other hand, the infinite projected entangled-pair states (iPEPS) algorithm \cite{peps_Cirac2004, Murg2007, Vidal2009, orus2009}, has also been adopted for many other 2D quantum lattice systems while computing the ground-state properties in the thermodynamic limit.
The implementation has been tested to work very well for the Heisenberg antiferromagnet \cite{Xiang2008, Wen2008}, hard-core bosons in a two-dimensional optical lattice \cite{Murg2007}, the quantum orbital compass model \cite{Oruscompass2009} and some frustrated spin-systems \cite{Murg2009} which definitely go beyond the reach of quantum Monte Carlo methods.
Following the precursors, An extended version of PEPS has also been developed to apply for fermionic systems with local interactions, called fPEPS \cite{fpeps_Cirac2010,fpeps_Vidal2010} where the sign problem can be circumvented.


In this work, we follow the TRG scheme to investigate some properties of the infinite temperature classical dimer model numerically and compare the results with conventional Monte Carlo (MC) techniques.
The relevant observable we choose to analyze the model is the dimer-dimer correlation function on a square lattice and a triangular lattice, being representative of the respective bipartite and non-bipartite class. 
In the former case, the correlation function falls off algebraically since the system becomes critical (at high temperatures), while in the later case, it develops a finite correlation length implying the existence of a liquid phase.
Our results show that for short range dimer correlations, tensor network representation performs very well with great accuracy even at low cut-off on the bond dimension.
However, for the critical phase, such as dimer model on square lattice, the TRG method is not very efficient due to rapid loss of entanglement related to the truncation of the bond dimension.


The paper is organized as follows.
We start by briefly reviewing the Pfaffian techniques used for exact calculations of the dimer correlations on square and triangular lattices in Sec. \ref{pfaffian}.
In Sec. \ref{method}, we present the numerical methods in two parts: in the first part, we demonstrate the implementation of the classical Monte Carlo method and in the second part, representing the weight of a configuration in terms of tensor network, we explain how to obtain the correlation functions using TRG.
In Sec. \ref{results}, we evaluate the dimer-dimer correlation by MC and TRG algorithm.
Discussing the behaviors of the correlation function on both the lattices we compare the numerical outcomes with the exact results obtained by the Pfaffian technique. 
We conclude in Sec. \ref{Conclusion} by giving a short summary on the applicability of TRG approach for studying gapped models and commenting on its performance particularly for the critical systems.

\section{Review of Pfaffian method}
\label{pfaffian}

The idea behind the implementation of the Pfaffian method essentially addresses two important aspects of the problem, i) exact computation of the partition function $Z$ in terms of the Grassmann fields and ii) calculation of the correlation functions (e.g. dimer-dimer) in the form of Green functions.
We divide this section in two parts accordingly.

\begin{figure}[ht]
\center{\epsfig{figure=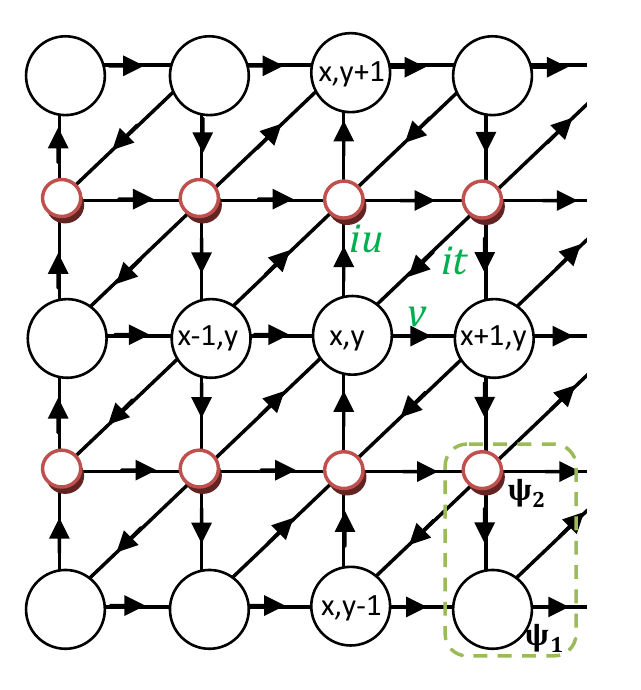,angle=0,width=5cm}}
\caption[]{The square lattice with diagonal bonds and a two-site (black and red circles) unit cell. Also shown are the directions of the Kasteleyn arrows which satisfy the clockwise odd rule described in the text.}
\label{latt}
\end{figure}

\subsection{Partition function}
\label{PF}

The partition function of the problem counts the number of dimer coverings on the given planar lattice \cite{ Kasteleyn1961, Fisher1961, Fendley2002,Wang2007,Loh2008,John2008}.
In order to construct the Pfaffian, we need to create an antisymmetric matrix $M$, called Kasteleyn matrix, and the partition function can be  conveniently expressed as a linear combination of the Pfaffians of four different Kasteleyn matrices $M_{(a,b)}$  where $a,b = 0 (1)$ implies periodic (anti-periodic) boundary conditions along the horizontal and vertical directions.
The elements of the matrix $M$ are defined to satisfy Kasteleyn's clockwise-odd rule: the number of arrows pointing in the clockwise direction around the faces is always odd.
A matrix element $M_{ij}$ is nonzero only if sites $i$ and $j$ are connected by a bond.
If $M_{ij}=1$, the arrow points from $i$ to $j$ and if $M_{ij}=-1$, the arrow points from $j$ to $i$.
The number of dimer coverings ( or equivalently the partition function $Z$) is given by $Z=| Pf [M] |$, where $Pf [M] $ is the Pfaffian of matrix $M$ where: $ Pf [M]= (\det [M])^{1/2}$.
The determinant can be calculated by using Fourier transformation of a fermionic path integral.
A fermion on site $i$ is associated with a Grassmann variable $\psi_i$, and the action can be defined as $S=\sum_{i<j}M_{ij}\psi_i\psi_j$.
The partition function for this action reads as
\begin{align}
Z = \int [D\psi] \exp(S).
\end{align}

As described in Ref. \cite{Fendley2002} a triangular lattice can be viewed as square lattice if the diagonal bonds are included with fugacity $t=1$.
The unit cell, in this case, has to contain two sites, and in the present context it is chosen to be doubled in the vertical direction as shown in Fig.\ref{latt}.

A unit cell is associated with the position vector $R_i$ with an index $\alpha=1,2$ locating the two sites within the unit cell.
The fugacities $u$, $v$, and $t$ are assigned to each of the vertical, horizontal, and diagonal bonds respectively.
The two fermions in the unit cell at $R_i$ are denoted as $\psi_{\alpha,R_i}$ with $\alpha=1,2$.
The action in terms of the Grassmann variables $\psi$ can be written as
\begin{align}
S = \frac{1}{2} \sum_{R_i,\alpha} \sum_{R_j,\beta} M^{\alpha\beta}_{R_iR_j} \psi_{\alpha,R_i} \psi_{\beta,R_j}
\end{align}
with $M^{\alpha\beta}_{R_iR_j}=-M^{\beta\alpha}_{R_jR_i}$.
Periodic boundary enforces the condition: $M^{\alpha\beta}_{R_iR_j} = M^{\alpha\beta} (R_i-R_j)$, and the elements of matrix $M$ multiplied by proper weight are given by
\begin{align}
& M^{12}(0)=M^{21}(\hat{y})=iu \\ \notag
& M^{11}(\hat{x})=M^{22}(\hat{x})= v \\ \notag
& M^{21}(\hat{x}+\hat{y})= - M^{12}(\hat{x})=it.
\end{align}

Fourier transforming the Grassmann variables by the relation
\begin{align} \label{FTforpsi}
\tilde{\psi}_{\alpha, \vec{k}} = \sum_{R_i} e^{i\vec{k}\cdot R_i}  \psi_{\alpha,R_i},
\end{align}
the partition function is expressed as $Z = \int [D\psi] \exp(S) = Pf[M]$ with
\begin{align}\label{action}
S = \frac{1}{2}\sum _{\vec{k},\alpha,\beta} \tilde{\psi}_{\alpha, \vec{k}}  \tilde{M}_{\vec{k}}^{\alpha,\beta}   \tilde{\psi}_{\beta,-\vec{k}}
\end{align}
where $\tilde{M}_{\vec{k}}$ is a $2\times2$ matrix:

\begin{align}
 \tilde{M}_{\vec{k}} =\left(
                      \begin{array}{cc}
    2iv\sin k_x & g(\vec{k}) \\
    g^*(\vec{k}) & 2iv\sin k_x \\
                      \end{array}
                    \right)
\end{align}
with $g(\vec{k}) =i[u-te^{ik_x}-ue^{-ik_y}-te^{-i(k_x+k_y)}]$ and $g(\vec{k})= -g^*(-\vec{k})$.

It is straightforward to calculate the partition function which follows from $Z^2=\det[M]$ where
\begin{align}
\det [M] = \prod_{\vec{k}}\det\tilde{M}_{\vec{k}},
\end{align}
and $\det\tilde{M}_{\vec{k}} = -4v^2\sin^2k_x-4u^2\sin^2(k_y/2)-4t^2\cos^2(k_x+k_y/2)$.

Subject to the periodic boundary conditions, the Pfaffians are the square root of the determinants specified by the Kasteleyn matrices  of lattice edges with, or without, the reversal of arrows on edges connecting two opposite boundaries.
A closer inspection of the Kasteleyn matrices $M(a,b)$ with proper boundary conditions reveals that the boundary Kasteleyn matrices will depend on the boundary conditions with phase factors along the horizontal and vertical direction.
The number of total dimer coverings with periodic boundary conditions is, thus, given by
\begin{align}
\label{Z_function}
Z = \frac{1}{2}(  & -Pf[M(0,0)]+ Pf[M(0,1)] \\ \notag
                  &+Pf[M(1,0)] +Pf[M(1,1)] ).
\end{align}
where the parameter $\kappa$ is decided by the parity of the dimer configurations (for details refer to \cite{ Kasteleyn1961}).
The magnitude of the wave vectors $\vec{k}$ depends on the boundary conditions $a$ and $b$ of lattice $L_x \times L_y$, such as $ k_x = 2\pi (l_x+a/2)/L_x $ and $ k_y = 2\pi (l_y+b/2)/L_y$ where $l_x=0,1,\ldots, L_x-1$ and  $l_y=0,1,\ldots, L_y-1$.

\subsection{Green function }
\label{GF}

Following  Ref.\cite{Fendley2002}, it is convenient to express the correlation functions in terms of the two-point Green function of the Grassmann variables,
\begin{align}
G_{ij}\equiv \langle \psi_i \psi_j\rangle= \frac{1}{Z}\int[D\psi]  \psi_i \psi_j \exp(S)
\label{green}
\end{align}

For any dimer covering of the lattice, the bond is either occupied by a dimer or not, so the number of dimers on each bond can only take the value $0$ or $1$.
Because the unit cell at most contains only one dimer, the probability of finding a dimer on a bond oriented along $\hat{y}$ in a unit cell at $R_i$ is given by
\begin{align}
\mathcal{P}(R_i) = | \langle \psi_{1,R_i} \psi_{2,R_i} \rangle|,
\end{align}
which is averaged over all positions on the lattice.
In a similar way, the dimer-dimer correlation function can also be expressed in terms of the Grassmann variables,
\begin{align}
D(r) = \langle \psi_{1,R_i} \psi_{2,R_i}  \psi_{1,R_j} \psi_{2,R_j} \rangle,
\end{align}
where we consider a case of $R_j=R_i+r\hat{x}$.

A Wick decomposition would cast the above form into products of Green functions. With the definition in Eq.(\ref{green}), it becomes
\begin{align}
D(r) = & \langle \psi_{1,R_i} \psi_{2,R_i}   \rangle   \langle \psi_{1,R_j} \psi_{2,R_j}   \rangle \notag \\
     - & \langle \psi_{1,R_i} \psi_{1,R_j}   \rangle  \langle \psi_{2,R_i} \psi_{2,R_j}  \rangle  \notag  \\
     + & \langle  \psi_{1,R_i} \psi_{2,R_j}  \rangle \langle \psi_{2,R_i} \psi_{1,R_j}   \rangle.
\end{align}

The Green function elements can also be read from the matrix $M$ using
\begin{align}
G_{ij}\equiv \langle \psi_i \psi_j\rangle = (M^{-1})_{ji} =  - (M^{-1})_{ij},
\end{align}
By using Eq.(\ref{FTforpsi}), the Green function can be written in the Fourier basis as $G_{R_i,\alpha;R_j,\beta}= \langle \psi_{\alpha,R_i} \psi_{\beta,R_j}\rangle = \sum_{\vec{k_i},\vec{k_j}} e^{-i(\vec{k_i}R_i+\vec{k_j}R_j)} \langle \tilde{\psi}_{R_i, \vec{k_i}}  \tilde{\psi}_{ R_j,\vec{k_j}} \rangle$ where we know $ \langle \tilde{\psi}_{R_i, \vec{k_i}}  \tilde{\psi}_{ R_j,\vec{k_j}} \rangle = \delta_{\vec{k_j},-\vec{k_i}} (\tilde{M}^{-1})^{ \alpha,\beta}_{\vec{k_i}}$ from Eq.(\ref{action}).
Finally, the two-point function can be represented in terms of $g(\vec{k})$ and  $\Delta(\vec{k})$ in the momentum space, namely,
\begin{align}
&\langle \tilde{\psi}_{1, \vec{k}}  \tilde{\psi}_{ 1,-\vec{k}} \rangle =  \langle \tilde{\psi}_{2, \vec{k}}  \tilde{\psi}_{ 2,-\vec{k}} \rangle = \frac{2i \sin(k_x) }{ \Delta(\vec{k})} \notag  \\
&\langle \tilde{\psi}_{1, \vec{k}}  \tilde{\psi}_{ 2,-\vec{k}} \rangle = - \frac{g^*(\vec{k})  }{ \Delta(\vec{k})} \notag
\\
&\langle \tilde{\psi}_{2, \vec{k}}  \tilde{\psi}_{1,-\vec{k}} \rangle = - \frac{g(\vec{k})  }{ \Delta(\vec{k})}
\end{align}
By doing a further Fourier transform, we can determine the real space Green function,
\begin{align}
\langle \psi_{\alpha,R_i} \psi_{ \beta,R_j} \rangle & = \int d\vec{k} e^{-i \vec{k} \cdot (R_i-R_j)} \langle \tilde{\psi}_{\alpha, \vec{k}}  \tilde{\psi}_{\beta,-\vec{k}} \rangle \\ \notag
 & = \int_0^{2\pi} dk_x\int_0^{2\pi} dk_y e^{-i \vec{k} \cdot (R_i-R_j)} \langle \tilde{\psi}_{\alpha, \vec{k}}  \tilde{\psi}_{\beta,-\vec{k}} \rangle \\
\end{align}

With a setting $u=v=1$, the fugacity $t=0$ gives us the square lattice and $t=1$ reproduces the triangular lattice.
Below we list some values of the dimer-dimer correlation function which are extracted by numerically integrating the above functions,
\begin{center}
\begin{tabular}{l c c  r }
\hline  $r$  \vline & $D(r)_{t=0}$ ~~~ \vline & $D(r)_{t=1}$ \\
 \hline 
   1 \vline & 0.12505080 \vline & 0.0466891518 \\
   2 \vline & 0.05780328 \vline & 0.0268276742 \\
   3 \vline & 0.07535259 \vline & 0.0281878353 \\
   4 \vline & 0.06192191 \vline & 0.0277736735 \\
   5 \vline & 0.06695547 \vline & 0.0277821043 \\
   6 \vline & 0.06235789 \vline & 0.0277783224 \\
   7 \vline & 0.06465416 \vline & 0.0277778197 \\
   8 \vline & 0.06242982 \vline & 0.0277777879 \\
   9 \vline & 0.06375758 \vline & 0.0277777821 \\
 \hline
\end{tabular}
\end{center}

\section{Numerical Methods}
\label{method}

\begin{figure}[ht]
\center{\epsfig{figure=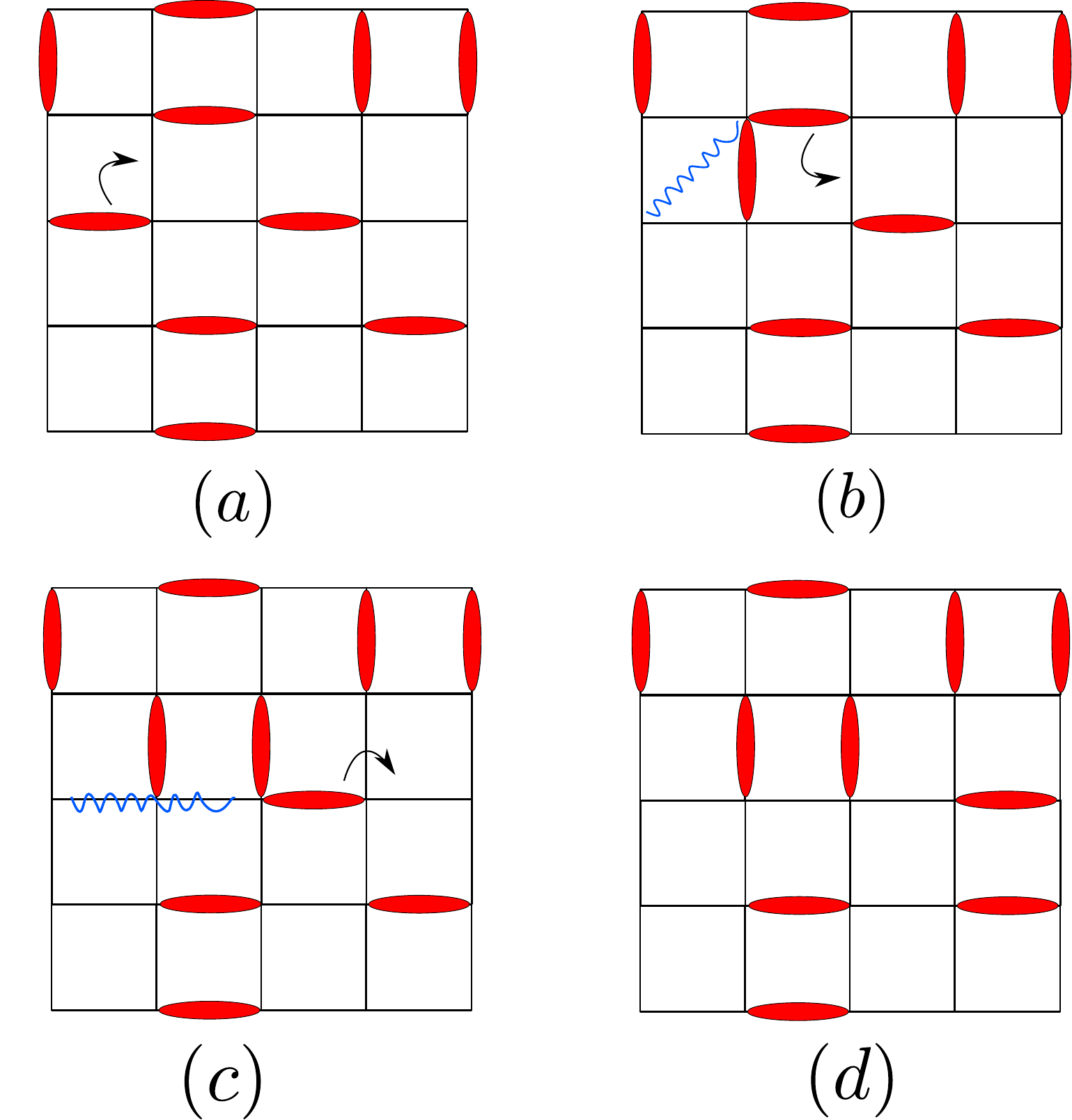,angle=0,width=6cm}}
\caption[]{(a) Dimer covering on square lattice. The arrow denotes the first move taken in Monte Carlo. (b) Two defects, joined by the string, are created at the sites having no dimer and two dimers. (c) The defects move further apart. (d) The final configuration achieved as consequence of the successive moves becomes a valid dimer configuration when the two defects merge. The arrows indicate successive dimer moves.}
   \label{MC}
\end{figure}

Since we are interested in studying the dimer models numerically, we deploy two different simulation schemes, TRG and classical MC, for comparing the results for clusters of different sizes on triangular and square lattice. 
As pointed out before, the standard MC algorithms have been very useful in simulating many of the classical problems by stochastically sampling the configurations based on Markov chain process \cite{newman1999monte}. 
We first present a brief account on implementing the MC algorithm for dimer models, then the same for the TRG construction. 
The details of the lattices are not very important for MC at the level of building the moves obeying two necessary conditions: detailed balance and ergodicity.
The algorithm is knows as ``long-loop worm algorithm'' which is known to satisfy both the conditions even for large system sizes \cite{newman1999monte}.

\subsection{Classical MC algorithm for dimer models}

On a square lattice of extent $L$, the total number of dimers is $L^{2}/2$ which corresponds to a filling fraction of 1/4 (because the number of links is $2L^{2}$).
We choose an arbitrary dimer configuration satisfying the hard-core constraint (no two dimers can meet at a site) to start the Markov process. 
In the first move, a dimer is randomly chosen and flipped to one of its six empty neighboring bonds as shown in Fig.\ref{MC}(a).
This generates two defects (we call them head and tail of the worm for demonstration) one at a site with no dimer attached (tail) and the other at the site with two touching dimers (head) (Fig.\ref{MC}(b)).
In the next step (Fig.\ref{MC}(c)) we move that dimer attached to the head which was present prior to the new dimer flipping in such that the defected head (with two dimers) always hops to a new site (we disdain the moves where it stays back even if the dimer flips to one of the empty bonds emanating from the head itself, e.g. if in Fig.\ref{MC}(b) the dimer on the horizontal bond attached to the head, denoted by the arrow, flips vertically).
This way we continue the successive moves until the defects meet each other and coincide in the very last move. This happens when one dimer always comes to fill one of the links attached to the fully emptied site generating a new valid configuration (Fig.\ref{MC}(d)).
The detailed balance is clearly satisfied as we always move to a new configuration (say, $\mu$) from an old one (say, $\nu$) with a finite probability $\mathcal{P}_{\nu\rightarrow\mu}$ where $\mathcal{P}_{\nu\rightarrow\mu} \propto \frac{1}{N}\cdot\frac{1}{6}\cdot(\frac{1}{3})^{l-1}$ in a system of $N$ dimers, hence all the states can be reached in finite time.
Note that $l$ is the total number of moves to connect two configurations $\mu$ and $\nu$.
In order to calculate the dimer-dimer correlations efficiently we construct a correlation matrix and average over all possible dimer locations along both the directions in a given configuration keeping the distance fixed.
The final correlation is then averaged over $\sim10^{10}$ configurations in lattices up to $L=128$.

%
%

Similarly, on a triangular lattice of size $L\times L$, the average dimer density corresponds to filling fraction of 1/6. The entire procedure carries over $mutatis-mutandis$ except, $\mathcal{P}_{\nu\rightarrow\mu} \propto \frac{1}{N}\cdot\frac{1}{4}\cdot(\frac{1}{2})^{l-1}$ and the correlation is calculated along the horizontal direction with parallel dimer ordering only.

\subsection{Tensor network renormalization}

Tensor network renormalization technique is based on real space regrouping of tensors defined on the sites of the underlying lattice.  
Assigning numbers (0 or 1) to the indices of the tensors, in practice, specifies the configuration with a certain weight and contributes to the partition sum $Z$ by that fraction.
The correlation functions can be extracted numerically by relating to $Z$  which is computed by successive tensor contractions.

\subsubsection{Tensor network}
\label{dimerTPS}

\begin{figure}[ht]
\center{\epsfig{figure=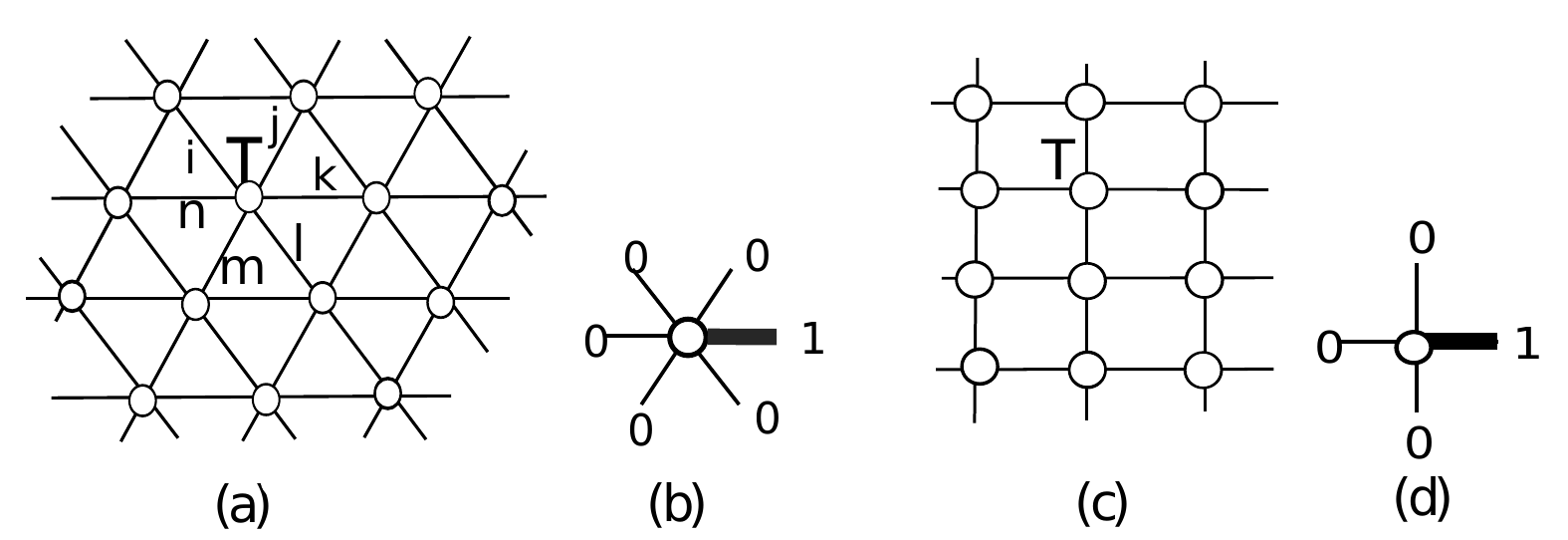,angle=0,width=9cm}}
\caption[]{ (a) and (c) Tensor network representation for 2D lattices.  Example of a dimer state (of a site) represented by tensor network on triangular lattice (b) and square lattice (d).}
\label{TPS-dimer}
\end{figure}

Configurations of the edges around each site in a valid dimer configuration can be formally described in tensor network representation $T_{ijkl...}$ with virtual indices $i,j,k,l,...=\{0,1\}$  on the triangular and square lattice as shown in Fig.\ref{TPS-dimer}(a-b) and (c-d) respectively.
The statistical weight for a given configuration can de measured as 
\begin{align}
W(i,j,k\cdots) = T_{ijklmn} T_{iopgrs} \cdots,
\end{align}
where the virtual bond includes $``0"$ and $``1"$.  The virtual index ``0"(``1") indicates absence(presence) of a dimer along the edge. 
For a local tensor, for example, $T_{ijklmn}$ on the triangular lattice, the hard-core constraint
\begin{equation}
T_{ijklmn} = 
\begin{dcases}
  1 & \text{if~~} (i+j+k+l+m+n) =1 ; \\
  0 & \text{otherwise,} 
\end{dcases}
\end{equation}
can be used to ensure each vertex to be connected to one dimer only.
The partition function is the sum of the weights of all possible configurations:
\begin{align}
Z = \sum_{i,j,k,l\cdots} W(i,j,k,l\cdots) = \text{tTr} (T_{ijklmn} T_{iopgrs} \cdots),
\end{align}
where ``tTr" means the tensor trace that all indices on connected links in the tensor network are summed over.

\begin{figure}[ht]
\center{\epsfig{figure=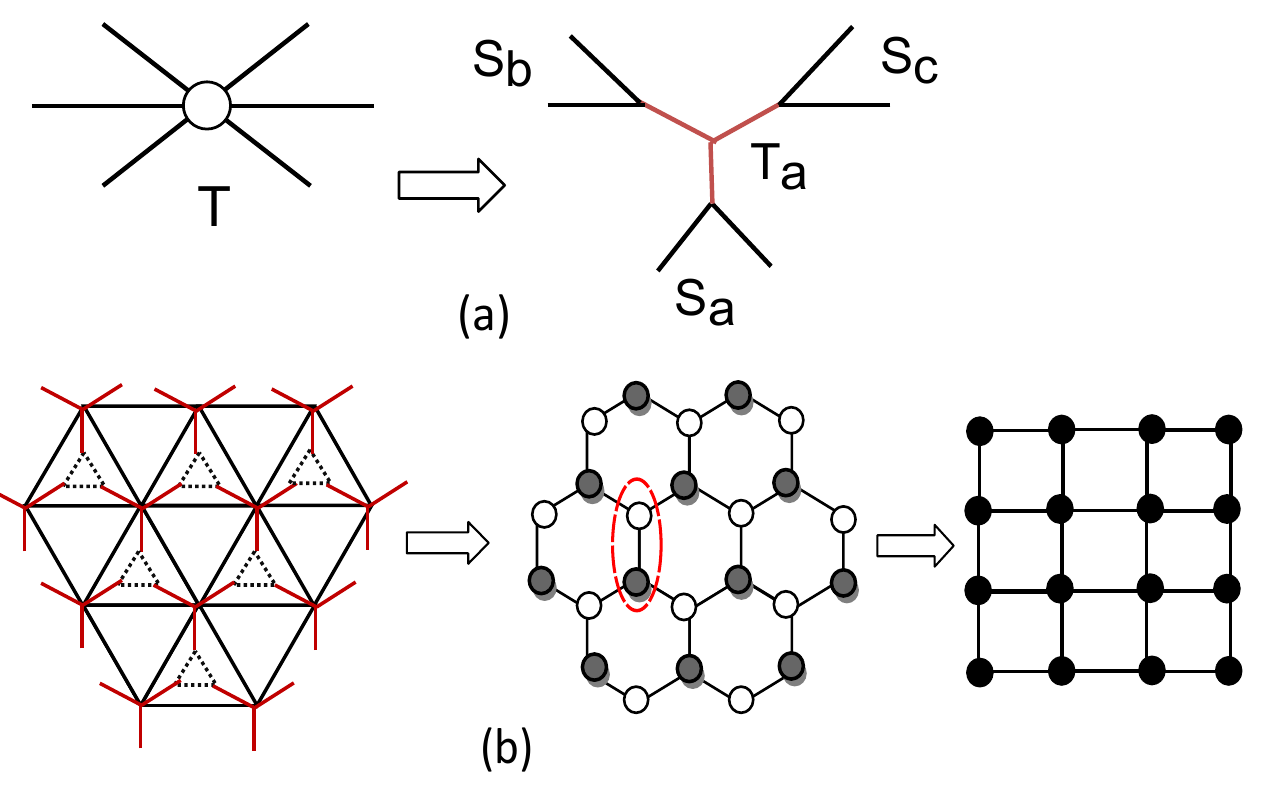,angle=0,width=8cm}} \caption[]{(a): Decomposing the tensor $T$ into $S_a$, $S_b$, $S_c$, and $T_a$. (b): Combining the three tensors in a dashed triangle $S_a$,$S_b$, and $S_c$ to form a new tensor $T_b$. The triangular lattice, thus, deforms into a hexagonal lattice with $T_a$ (white) and $T_b$ (black) tensors on the respective sublattices. Further grouping of the two tensors in every basis of the hexagonal lattice generates a new square lattice network of rank-4 tensors.}
\label{trg}
\end{figure}

\subsubsection{Renormalization algorithm}

%
%
In two-dimensional system, it is, however, difficult to calculate the tensor trace (tTr) since all indices on the connected links in the network need to be summed over.
This imposes the hurdles of an exponentially hard calculation.
Several approximation schemes have been proposed as solutions in this context such as iPEPS algorithm, the corner transfer matrix method (CTMRG) \cite{CTMRG-09}, and tensor renormalization approach \cite{trg07-Levin,trg08-Wen} which tackle this problem essentially by scaling the computational effort down to the polynomial level of calculating the tensor trace.
In this paper, we use the tensor renormalization approach which is akin to the real space renormalization in the way that at each step, the RG is structured by merging sites (by contracting respective tensors) and truncating the bond dimension according to the relevance of the eigenvalues in the Schmidt decomposition of the old tensors.
We resort to this technique for calculating the partition sum and describe the steps, for example, on a triangular lattice starting with the original local tensor $T$.
In the process, we first split it into four parts, $S_a$, $S_b$, $S_c$ and $T_a$ with several singular value decompositions as shown in Fig \ref{trg}(a), which, in turn, changes the lattice structure after the first step.
The second step is to build a new rank-three tensor denoted by $T_b$ (see Fig. \ref{trg}(b)) as follows:
\begin{align}
\label{trgstep1}
T_{b(\alpha\beta\gamma)}=\sum_{ijk}S_{a(ij\alpha)}\times
S_{b(jk\beta)} \times S_{c(ki\gamma)}.
\end{align}
The third step is to merge two sites to form a new rank-$4$ tensor shown in Fig. \ref{trg}(b). 
The triangular lattice tensor network is now mapped to a square model where applying TRG is known to be simple and straightforward \cite{Xiang2008}.
Each step of TRG reduces the number of sites by half.
Eventually, the entire network collapses to only a few sites and the double tensor trace appearing in the partition function $Z$ can be calculated easily.
%

\begin{figure}[ht]
\center{\epsfig{figure=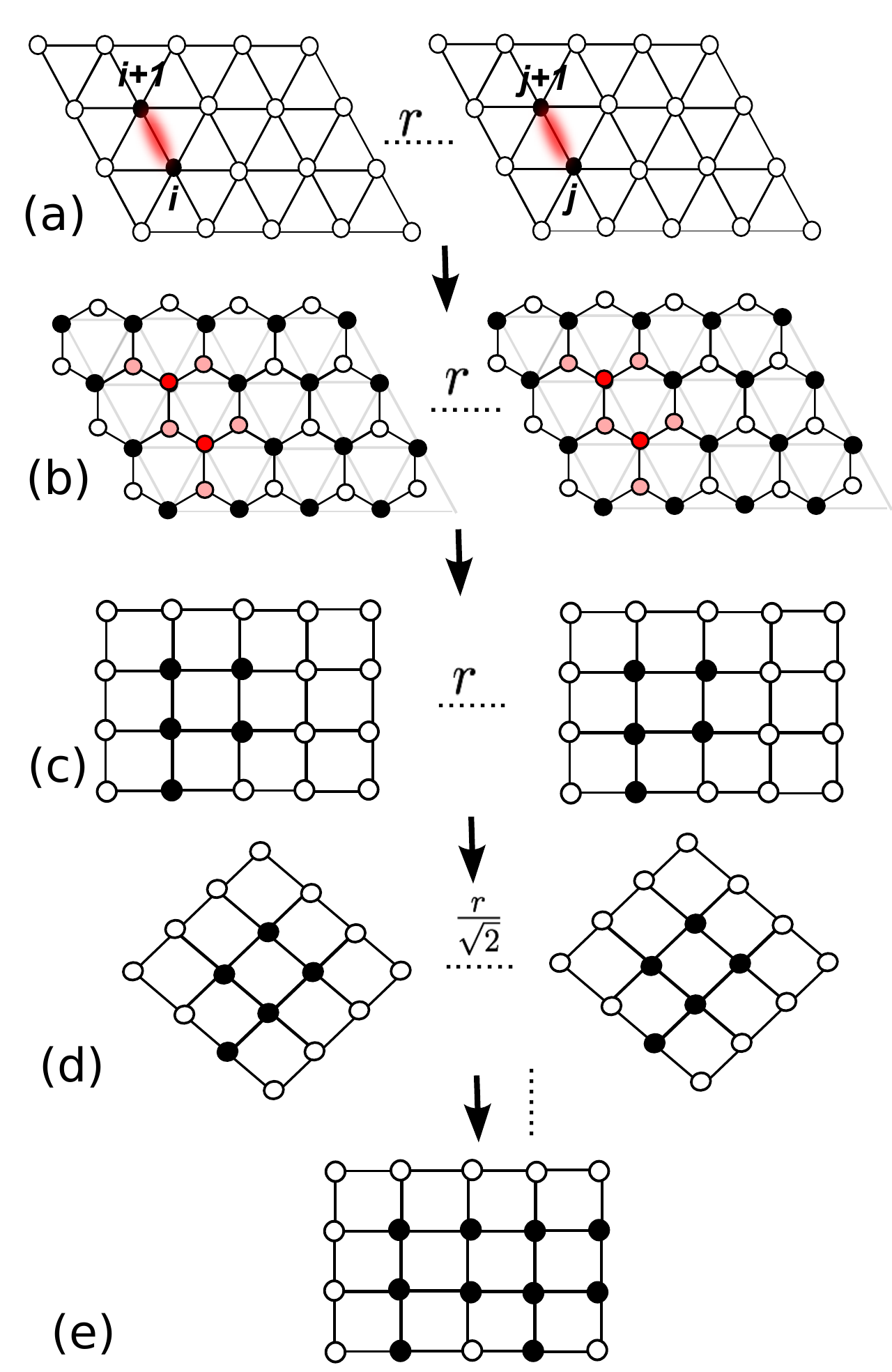,angle=0,width=8cm}} \caption[]{Schematic of the tensor renormalization for calculating dimer-dimer correlation function at a distance  $r$.  (a)(c)(d)(e):The black dots represent the impurity tensors and white dots represent the uniform tensors. (b): The red dots represent the impurity tensors while white (black) dots represent the uniform tensors on respective sublattices. }
\label{trg_corr}
\end{figure}

In the following, we briefly discuss how to calculate the dimer dimer correlation function in this tensor network. 
We assume the distance between two dimers to be $r=|R_i - R_j|$ and use the dimer counting operator  $d(R_i)$ for a given link $R_i$.
The correlation functions  $\langle d(R_i) d(R_j) \rangle$ can be represented as a tensor trace with four impurity tensors living on the sites $i,i+1$ and $j,j+1$ as shown in Fig. \ref{trg_corr} (a) and uniform tensors $T$ on every other sites. 
After the first step of TRG, we decompose these tensors and form the new rank-3 tensors as prescribed before which leads to a new network on a honeycomb lattice where the number of the impurity tensors ( red dots in Fig. \ref{trg_corr} (b)) increases to 14. 
By blocking every two tensors in one, we arrive at a square lattice network comprising of the new rank-4 tensors with the impurity tensors being 10 in number.
In the process of further coarse-graining, the lattice structure remains intact with the same number of impurity tensors as shown in Fig. \ref{trg_corr} (c)-(e), however the distance renormalizes by a factor of $1/\sqrt{2}$ in every step. 
%
%
Finally, we end up with a tensor network of the size we expect, and obtain the correlation function by performing the tensor trace on it. 

\begin{figure}[ht]
\center{\epsfig{figure=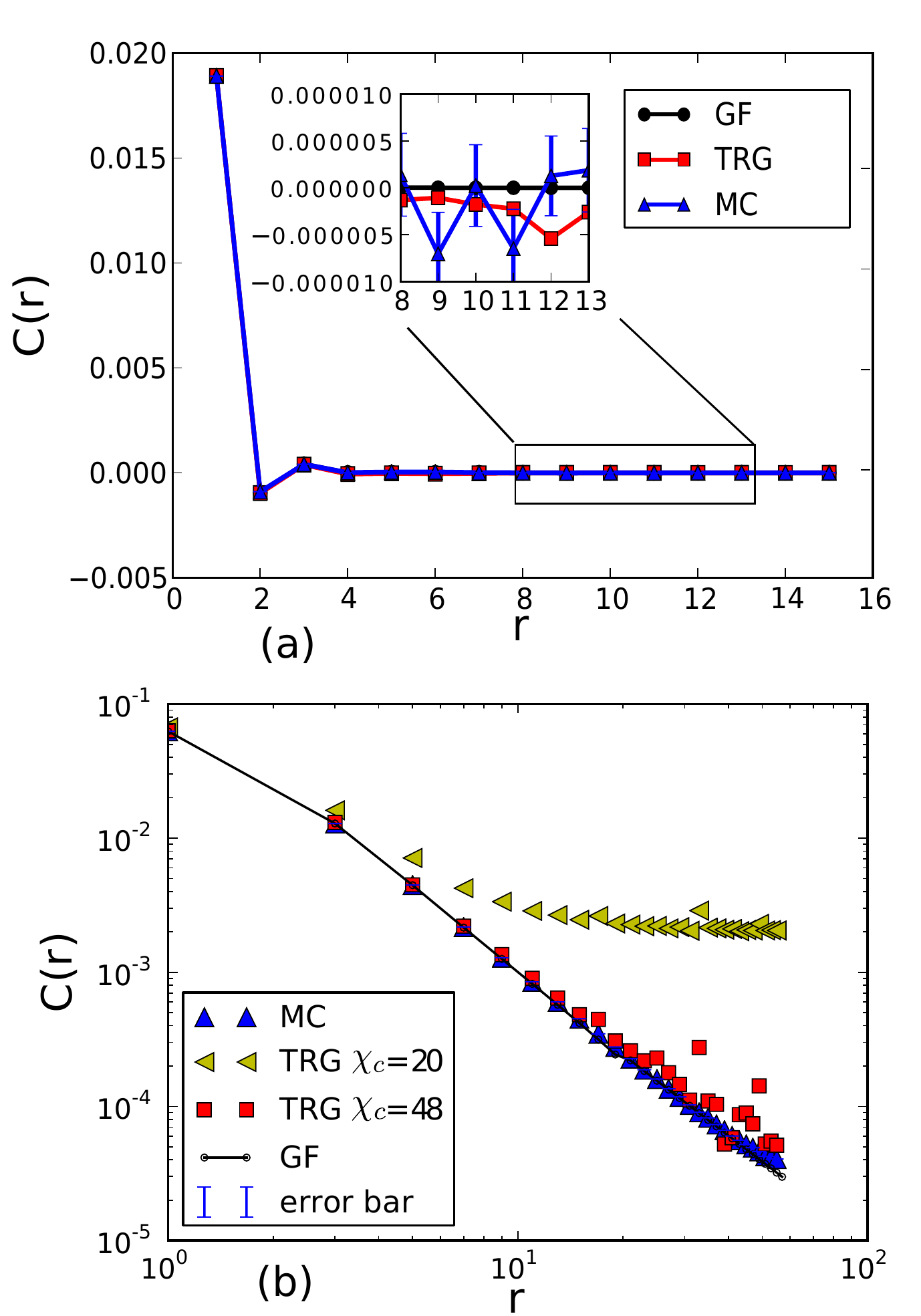,angle=0,width=9cm}}
\caption[]{
(Up): The dimer-dimer correlation on the triangular lattice as a function of the distance $r=|R_i-R_j|$. The plot shows data for an $128\times128$ cluster as extracted from  Green function, TRG method, and MC method. In the tensor network representation the bond dimension has been truncated at $\chi_{c}=20$.
 (Down):The same function is plotted for a $128\times128$ square lattice cluster using Green function, MC method and TRG method with $\chi_c=20, 48$. That the cut-off is kept very high for required accuracy is attributed to the critical nature of the model on square lattice.
  }
   \label{dodr}
\end{figure}

\begin{figure}[ht]
\center{\epsfig{figure=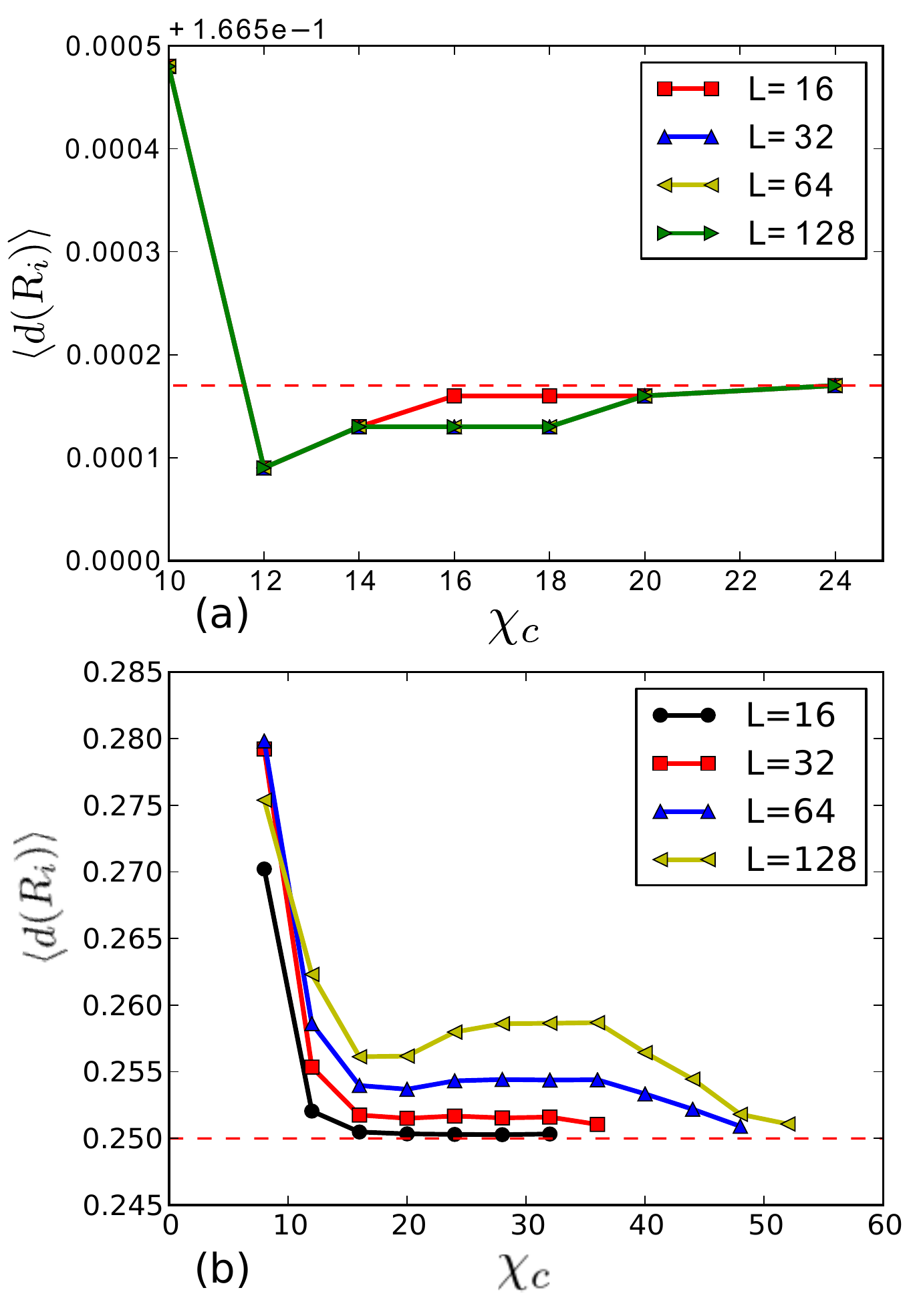,angle=0,width=8cm}}
\caption[]{ 
(Up): The dimer concentration as a function of the cut-off for different lattice sizes on the triangular lattice. The red dotted line indicates the average value of $1/6$.
(Down): Shown is the above quantity for the square lattice. The average value in this case is $1/4$. Note the deviations from the dashed line for larger systems. The convergence for them requires larger cut-off which grows with the system size.}
   \label{one_dr}
\end{figure}

\section{Results}
 \label{results}

The relevant quantity of our interest is the dimer-dimer correlation which appears very useful in characterizing different phases of the classical dimer models on bipartite and non-bipartite lattices. 
%
%
%
The connected correlation function is expressed as,   
\begin{align}
C(r)=\langle d(R_i) d(R_j) \rangle-\langle d(R_i)\rangle \langle d(R_j),
\rangle.
\end{align}
where $r=|R_i-R_j|$ and $d(R_i)$ counts 1 if a dimer is present on the link $R_i$ in a given configuration otherwise 0.
Here the notation $\langle\cdots\rangle$ denotes the statistical average over all the configurations.
In the asymptotic limit of large $r$, the correlation converges to the square of the average dimer concentration $\langle d(R_i)\rangle$ which is 1/4 and 1/6 on the square and triangular lattice respectively.

In the MC simulation the above function is calculated for different clusters of size up to 128$\times$128 for both the lattices. 
The results are in well agreement with the Green function (GF) calculations and indicative for a finite correlation length $\sim$ one lattice spacing in case of the triangular lattice (Fig.\ref{dodr} upper panel) dimer model.
The error bars are very tiny since the averaging is performed over large number of statistical ensembles.
On the other hand, the dimer-dimer correlation on the square lattice is visibly algebraic of the known form of $|C(r)|\sim r^{-2}$ for large $r$ as displayed in the lower panel of Fig.\ref{dodr}.

%

Compared to the MC results, TRG also shows similar functional form of the correlator for the dimer model on the triangular lattice.
The TRG results (Fig.\ref{dodr} upper panel) are consistent even at large system sizes (up to 128$\times$128) and achievable at the cost of reasonable cut-off $\chi_c$.
In our case, the required convergence is obtained for $\chi_c=20$ which certifies TRG as an extremely useful machinery for studying gapped systems. 
However, for the square lattice dimer model, TRG turns out not so effective on account of large deviations from the exact results and MC while keeping the parameters (system size and $\chi_c$) same as the triangular lattice.
Even if we crank up the cut-off to $\chi_c=48$, unstable values appear in the correlation at large distances.
As it turned out, the convergence demands unusually high cut-off on $\chi$ for such large system sizes costing a huge amount of computational resources.

In order to provide a critical assessment about the performance of TRG for simulating dimer models on different lattices, we calculate, for example, another quantity called average dimer concentration $\langle d(R_i)\rangle$ as a function of $\chi_c$. 
From the results of TRG method shown in Fig.\ref{one_dr} upper panel, we observe that $\langle d(R_i)\rangle$ rapidly converges to 1/6 for the dimer model on the triangular lattice as the cut-off $\chi_c$ is gradually increased.
This is evidently not the case for the dimer model on the square lattice (Fig.\ref{one_dr} lower panel).
Even for small system sizes, the required stability in the numerical data requires cut-off like $\chi_c\sim30$.
If we go to larger system sizes, more and more relevant information has to be stored in a pertinent basis since the convergence is attained at a much larger values of $\chi_c$.
The failure can be attributed to the fact that we truncate the sum over indices as in Eq.~\ref{trgstep1} at a threshold much below the required one, so the entanglement dies off gradually.

\section{Conclusion}
 \label{Conclusion}

In this paper, classical dimer models on square and triangular lattice are studied numerically by using MC and TRG techniques.
We revisit the Pfaffian construction which provides an analytic way to tackle the statistical problem by computing the partition sum exactly.
Following the Green function formulation, one can obtain the expression for the dimer-dimer correlation function which displays very different behavior on the square and the triangular lattice hinting at a critical and liquid phase respectively.
We present the detailed structure of the simulations adopted here, particularly focusing on TRG, and trace the functional behavior of the dimer-dimer correlator numerically.
Our study conclusively establishes that TRG can serve as a very efficient and elegant platform to compute physical quantities of a gapped system such as dimer model on the non-bipartite lattices.
However, the performance and accuracy fall off significantly while capturing the physics of a critical phase i.e. dimer model on the square lattice.
The convergence requires large critical bond dimension $\chi_c$ for the tensor contraction if we want to take care of the consistency of the basis for large system sizes.
Entanglement in the network in this case diminishes very rapidly which raises the cut-off very high for storing the necessary information even at the minimal level and makes the simulation run beyond the capacity of our present resources.  
From the present work on the dimer models, it can be envisaged that TRG should be very useful for investigating classical loop gases on non-bipartite lattices e.g. fully packed loop model on the triangular lattice which has not been attempted so far.

\section*{Acknowledgments}
We would like to acknowledge Frank Pollmann for many sparkling discussions regarding the project and critically reviewing the manuscript. 


\appendix

\bibliography{bibs_jessie}

 \end{document}